\documentclass[12pt,preprint]{aastex}
\shortauthors{Hawley et al.} 
\shorttitle{Flares on YZ CMi}
\begin{document}

\title{Near-Ultraviolet Spectra of Flares on YZ CMi}

\author{Suzanne L. Hawley\altaffilmark{1},
Lucianne M. Walkowicz\altaffilmark{1},
Joel C. Allred\altaffilmark{2},
Jeff A. Valenti\altaffilmark{3}
\altaffiltext{1}{Astronomy Department, University of Washington,
   Box 351580, Seattle, WA  98195}
\altaffiltext{2}{Physics Department, Drexel University, 3141 Chestnut Street, 
Philadelphia, PA 19104}
\altaffiltext{3}{Space Telescope Science Institute, 3700 San Martin
Drive, Baltimore, MD 21218}}


\begin{abstract}

Near-ultraviolet spectroscopic data obtained with the HST STIS 
instrument on the dMe flare star YZ Canis Minoris (YZ CMi) 
were analyzed.  Flare and quiet intervals were identified from the
broadband near-UV light curve, and the spectrum of each flare
was separately extracted.  Mg II and Fe II line profiles
show similar behavior during the flares.  Two large flares
allowed time-resolved spectra to be analyzed, revealing a
very broad component to the Mg II k line profile in at
least one flare spectrum (F9b).  
If interpreted as a velocity, this component requires chromospheric
material to be moving with FWHM $\sim$ 250 km sec$^{-1}$, implying kinetic
energy far in excess of the radiative energy.  The Mg II k
flare line profiles were compared to recent radiative hydrodynamic
models of flare atmospheres undergoing electron beam heating.  
The models successfully predict red enhancements
in the line profile with typical velocity of a few km sec$^{-1}$, but
do not reproduce the flares showing blue enhancements, or
the strongly broadened line observed in flare F9b.  A more complete
calculation of redistribution into the line wings,
including the effect of collisions with the electron beam, may
resolve the origin of the excess line broadening.
\end{abstract}



\section{Introduction}

Spectroscopic signatures of flares on M dwarfs have been sporadically
explored, mostly at optical wavelengths, since the pioneering 
observations
of UV Ceti by Joy \& Humason (1949).  Principal features observed
during flares include strong continuum radiation rising toward
the blue and near-ultraviolet; broad, enhanced hydrogen Balmer series 
emission lines;
enhanced emission in high temperature lines such as He I, He II,
C IV, N V, Si IV; enhanced Xray emission; and bursts of radio emission
(e.g. Kunkel 1967, Jackson et al. 1989, Hawley \& Pettersen 1991,
Eason et al. 1992, Osten et al. 2005).

YZ Canis Minoris (YZ CMi, also known as Gliese 285), an active dM4.5e 
star with V=11.1 and d=6pc (Perryman et al. 1997), has been a popular 
target for flare 
monitoring campaigns, due to its proximity and strong
activity, which improves the chances of observing a flare. 
Extensive multiwavelength spectroscopic flare studies of YZ CMi 
include those of Kahler et al. (1982), Worden et al. (1984), and 
van den Oord et al. (1996).

Previously, no M dwarf flare star has been systematically observed
with high-resolution near-ultraviolet spectroscopy in the region 
where the important chromospheric Mg II h and k lines,
and numerous Fe II emission lines, are produced.  IUE did not
have the sensitivity to provide much time or spectral resolution
during flares, although strong Mg II emission was observed from the 
dM3e star AD Leo during the decay phase of an exceptionally large flare
(Hawley \& Pettersen 1991).  Linsky et al. (1982) carried out
an early pioneering study of ultraviolet emission in active cool stars 
with IUE, while a later
comparative study of Mg II h and k emission with X-ray emission
was undertaken for a sample of M dwarfs, including YZ CMi, by
Mathioudakis \& Doyle (1989).  Broadband near-ultraviolet
emission during flares has been recently studied, using XMM-Newton data 
by Mitra-Kraev et al. (2005), and using GALEX data by 
Robinson et al. (2005).

The data presented here were obtained
as part of a larger multiwavelength
campaign to observe YZ CMi in 2000.  We use these near-ultraviolet
spectra to examine the Mg II and Fe II emission line profiles 
in YZ CMi during quiescence and flares, and to study the 
behavior of the Mg II emission in the flares that are large
enough to be subdivided into separate, time-resolved flare spectra.  
We also compare the observational data for these chromospheric
lines with predictions from our new generation of radiative
hydrodynamic models (Allred et al. 2005, 2006).
The new models incorporate detailed radiative transfer
in atmospheres heated by non-thermal electrons and undergoing
significant mass motions as heated material flows upward
(evaporation) and downward (condensation) within a magnetic
loop.  The models were designed to predict velocities and
line profiles of chromospheric emission
lines, and are therefore suitable for comparison with the Mg II
observations described here.

\section{Observations}

The Hubble Space Telescope (HST) observed YZ CMi on 2000
February 5-6, as part of GO program 8129 (PI: R. Robinson).
We reanalyze these archival data as part of AR program 10312
(PI: S. Hawley). The Space 
Telescope Imaging Spectrograph (STIS, Kimball et al. 1998),
was used with the medium resolution NUV echelle grating 
(E230M) and the 0.2x0.2 arcsec aperture.  
The E230M spectra have a nearly constant ($\pm 1.5$\%)
dispersion of 4.88 km/s per pixel. The line spread function
(LSF) for the 0.2x0.2 arcsec aperture has a FWHM of 1.8-1.9 pixels
over the wavelength range 2400-2800\AA, corresponding to a
resolving power of R=33,000.
These are the only 
STIS spectra of an M dwarf flare star at NUV wavelengths that 
exist in the HST archive. Data were obtained during
six consecutive orbits, with exposure times ranging 
from 32 to 40 minutes,
over a 6.5 hour total timespan. The MAMA detector timetag
mode was used for the observations, resulting in an event
list for each exposure that contains the detection 
time and location of every detected photon.

To construct light curves of the integrated light obtained
during each exposure, we
used an extraction window that was 11 pixels wide, centered
on each echelle order, to assign all timetag events to either
source
or background regions on the MAMA detector. We then divided
each exposure into 100 uniform time bins (19-24 s in duration
depending on the individual exposure lengths which range from
32-40 min, see Table 1),
obtaining separate light curves for the source and background
regions. For each exposure, we fitted the background light
curve with a sixth order polynomial, scaled the result by
the ratio of source to background pixels, and then subtracted
this predicted background from the source light curve to obtain
a total observed stellar count rate for each time bin. Errors
for each time bin include Poisson noise from the subtracted
background.

Figure 1 illustrates the resulting light curves for
the six exposures.  The flaring and quiet intervals F1-10 and 
Q1-8 were assigned by qualitative inspection of the light
curves, and are labeled on the figure.  
Table 1 provides relevant information about each spectral
interval.
Column 1 lists the number of
the exposure containing the time interval. Exposure number N 
maps to HST dataset o59k010N0. The second and third columns
give the start and end time of each interval, expressed in
minutes since the beginning of each exposure. Column 4 gives
the label assigned to each time interval, with Q labels
indicating quiet intervals and F labels indicating flare
intervals.  The numbers were assigned sequentially in time
for individual intervals.  Columns 5 and 6 give the total
count rate and standard deviation for each interval.  Although
the flaring and quiet intervals were assigned qualitatively,
it is clear from these data that the flaring intervals have
both higher count rates and larger scatter.  The exceptions
are the late decay phases of the two large flares, labeled
as F2c and F9c.  Since we wanted to investigate the evolution
of the emission lines in these late phases, we retain those
as flare intervals.  Also note that quiet interval Q7 has a
relatively high count rate and shows significant low level
variability in the light curve, but no obvious flares.  
We include it as a quiet interval after insuring that the 
Q7 spectrum shows no more than 1.2\% variation compared to
the summed quiet spectrum over all other intervals.  Since
Q7 represents almost one-third of the quiet exposure time,
including it increases the signal-to-noise of the summed quiet
spectrum.

We used the IRAF package STSDAS (version 3.3.1, 2005 March 31)
to split the timetag event lists into the quiet and flaring
time intervals and then to extract spectra for these intervals.
To split the event lists, we used the task 'inttag' (version
1.1, 2000 January 26). To extract the spectra, we used the
task 'calstis' (version 2.18, 2005 February 15) with the
best calibration reference files available on 2005 July 10.
The calibration process removed radial velocity shifts caused
by orbital motion of the spacecraft around the Earth and the
Earth around the Sun. Hence, every extracted spectrum is in
(the same) heliocentric reference frame.
Table 2 provides the UT and Julian Date (JD) at the start
of each exposure.

YZ CMi has a measured rotation velocity $v$ sin $i$ = 6.5 km sec$^{-1}$
(Delfosse et al. 1998).  With a nominal radius of 0.2 R$_{sun}$,
this corresponds to a minimum period of 1.6 days, considerably
longer than the total (consecutive) elapsed time of the HST exposures
($\sim$ 0.4 days).
The individual exposures comprise $<$ 2\% of the minimum period, indicating
negligible change in the location of a given active region during a single
exposure.  We do not see strong evidence for smooth (non-flaring) variability
within a given orbit as was found by Robinson et al. (1999) in their
near-ultraviolet photometric observations of YZ CMi using HST with the High
Speed Photometer instrument.  Our analysis of the STIS
spectrosopic data is not presently 
suitable for comparison with their flare frequency analysis, but
it would be interesting to reanalyze the MAMA timetag data in sum
over shorter time intervals to more rigorously identify individual flares.
However, this is beyond the scope of the present paper.

\subsection{Light Curves and Spectra}

Figure 1 and Table 1 show that YZ CMi was in an obvious 
flaring state for 
approximately 40\% of the total 224 minutes of exposure time comprising
the six separate exposures.
The flux emitted in the near-ultraviolet bandpass sampled
here (2300-3050\AA) is approximately 4.2 $\times 10^{-12}$ 
ergs sec$^{-1}$ cm$^{-2}$ 
during quiet periods.
Using a distance of 6 parsecs for YZ CMi, the quiescent energy emitted
from the visible hemisphere
during the 224 minutes of exposure was $\sim 1.2 \times 10^{32}$ ergs, 
while the energy emitted by flares was $\sim 1 \times 10^{31}$ ergs, 
giving a total energy in this bandpass of $\sim 1.3 \times 10^{32}$ ergs.
The flare energy comprises $\sim$ 8\% of the total.

Figure 2 illustrates the spectrum summed over all quiet intervals.
The line identifications were made using the Chianti database
(Dere et al. 1997, Landi et al. 2006).
Clearly most of the emission lines in the
near-UV come from Fe II, with the notable addition of the Mg II
h and k lines and minor contributions from Si II, S II and
Al II.

Figure 3 shows the summed spectrum over all flare intervals, with
the quiet spectrum subtracted. Note that subtracting the quiet spectrum
without scaling assumes that the area covered by the flare is 
negligible compared to the area of the visible hemisphere of the
star.  As discussed in Section 3, the flare area required
to best match the observed Mg II k emission to the model predictions 
is $\sim$ 0.5\% of the observed surface, which supports the assumption that
the flare area coverage is negligible.  The flare-only emission
shown in Figure 3 is primarily evident in the Mg II and Fe II lines,
which are enhanced by a similar amount ($\sim$ 10\%) over the quiet
spectrum.  Significant broadening in the Mg II lines is also
apparent in this flare-only spectrum, and is discussed further
in section 2.2 below.

Table 3 gives the energy breakdown between
lines and continuum for the flare intervals individually 
and in sum.  The first column shows the excess flare line flux calculated
from the subtracted spectrum (flare interval - quiet), with
the value in parentheses being the ratio of the flare line
flux to the total (line + continuum) flux.  The second
column shows the same calculation for the continuum.
Thus the values in parentheses represent the relative importance
of lines and continuum to the excess flare emission, and sum to 1
in each case.  As typically
seen in stellar flares (e.g. Hawley \& Pettersen 1991; Hawley et al.
2003) the larger flares are dominated by the continuum contribution, 
e.g. F2a, F4, F7 and F9a where $\sim$ 70-90\% of the energy is contained
in the continuum.  Two of the flares (F1 and F3), actually
showed negligible line enhancements (formally negative but
within the uncertainties of being small scale variability)
but well-detected continuum enhancements.
A large M dwarf flare was recently observed with GALEX with a broad
bandpass NUV filter (Robinson et al. 2005), and analyzed assuming 
that all of the NUV
emission was from the continuum.  This appears to be reasonably
justified by our analysis of the NUV flare spectra presented here.
However, some smaller flares (F5, F8), and the decay phases of large
flares (F2b, F2c, F9b), emit as much as 50\% of their energy in the 
emission lines.  In sum, the flaring intervals show that
most flare energy is emitted in the continuum (77\% compared to
23\% in line emission), while the quiet spectrum is dominated
by the emission lines, with 65\% of the energy emitted in 
lines compared to 35\% in the continuum.

The NUV emission line energy is quite 
evenly divided between Mg II and Fe II in this bandpass, with
the Fe II/Mg II energy ratio being 0.97 in the quiet spectrum, and
1.01 in the flare spectrum. Each Mg II line carries much more energy
than any individual Fe II line, but there are a large number (78)
of identified Fe II lines in this region, which makes them
energetically important in sum.  Our results for the quiet
ratio are in agreement with those found in Linsky et al. (1982).

\subsection{Flares}

The Mg II lines undergo significant changes in their line profiles
during the flaring intervals.
Figure 4 shows the Mg II k line profiles
for the individual flares (F1-10), obtained from the summed spectra over
the extent of each flaring episode.  Flares F2 and F9 were large enough
that it was possible to split them up into impulsive, peak and decay phases,
shown as F2abc and F9abc in Figure 4.  The quiet flare profile is shown
for comparison as the dotted line in each panel.  It is clear that 
the quiet flux dominates
the total even during flares, indicating that the flare area coverage
is very small, and that Mg II k is already a very strong line even in the
quiet chromosphere.  This behavior is similar to H$\alpha$, which
also responds only slightly to flares in early-mid M dwarfs (e.g.
Hawley et al. 2003), though it responds 
more strongly in late M dwarfs (cf. Liebert et al. 1999). This is
in contrast to lines such as He I and II, and the higher
order Balmer lines, which become significantly stronger during
flares, often dwarfing the quiet contribution to the line flux
(Hawley \& Pettersen 1991).

Figure 5 shows the subtracted profiles for each flare (i.e. flare - quiet,
as in Figure 3),
in order to examine possible velocity shifts during the flare.  The
flares are shown on the same vertical scale such that strong flares
have significant excess emission, and small flares have very little.
It is clear that some flares (F1, F2, F3, F7, F10) show a blue
wing enhancement and red wing deficit, while others show the
opposite behavior (red wing enhancement and blue wing deficit -
F4, F9) and still others show both blue and red enhancements
(F5, F8) or deficits (F6). 
For simplicity, we use only the Mg II k line 
in the following analysis, as the Mg II h line exhibited very 
similar behavior in all cases.

The flares are grouped according to the appearance of the excess flare
emission in Figure 6 (blue enhancements) and Figure 7 (red
enhancements), with the left panels showing Mg II k and the right
panels the strong Fe II UV1 line at 2600.2\AA.
In general, the enhancements
seen in Mg II k are also seen, albeit at a lower level and with poorer 
signal-to-noise ratio,
in the Fe II data, as expected since these lines are formed in
approximately the same temperature region of the atmosphere.
Simple Gaussian fits to the enhancement features give wavelength
shifts for each of these flares, as shown in Table 4.  The shifts
are not large, typically a few km/sec in both the upward (blue shift)
and downward (red shift) directions.  
The Sun also shows such velocity shifts even outside of obvious flares.
These have been attributed to asymmetric heating in active region loops
(Winebarger et al. 2002).

To verify that these apparent velocity shifts are not artifacts of the
data reduction, we compared the magnitude and sense of the shift 
in each flare with the peculiar velocities of HST and the Earth during 
the time period in the orbit that the flare occurred, and found 
no correlation.  We see no evidence that the observed velocity
shifts are instrumental in nature, and therefore presume that they
represent real effects in the NUV emission lines emitted by the star.
We discuss the velocities compared to those expected from our 
flare models in Section 3 below.

Figure 8 shows the time evolution of the Mg II k emission
for the largest flares, F2 and F9.  Flare F2 exhibits rather
simple behavior, with the narrow central profile strongly enhanced
on the blue side in the impulsive phase (F2a), and decreasing
monotonically during the flare decay (F2b, F2c). In contrast,
flare F9 is quite unusual, initially (F9a) showing a strong
red enhancement in the central profile, but then developing
very broad, symmetric wings in F9b. Upon closer inspection of
Figure 4, it appears that there is evidence for similarly
broad wings in F7 and F9a.  These data are reminiscent
of those in Doyle and Byrne (1987) where extensive, symmetric
line broadening was seen in the hydrogen Balmer lines
during a flare on YZ CMi.  Broad hydrogen lines during flares
are usually interpreted as being due to Stark broadening 
(Donati-Falchi et al. 1985, Hawley and Pettersen 1991, Johns-Krull 
et al. 1997, Jevremovic et al. 1998)
but Doyle and Byrne were unable
to find a satisfactory fit to the data with Stark profiles.
Instead they inferred symmetric red and blue shifted emitting
regions, or turbulent motions, of order a few hundred km sec$^{-1}$.
This is typical of the broad component often seen in transition region and
coronal emission lines which is often ascribed to overlapping emission
from multiple explosive microflaring events, leading to high velocity
mass motions (Antonucci et al. 1993, Wood et al. 1996, 
Wood, Linsky \& Ayres 1997, Vilhu et al. 1998).  However, those
broad components are not associated with large flares, but appear
even in the quiescent spectra of active (earlier-type) stars, and the Sun.
The phenomenon we have observed appears to be directly related
to the large flare heating event F9, and does not appear in quiescence
or in the other flaring events (except possibly flare F7).  In 
particular it does not occur in the other large flare we observed, 
flare F2.  

In the present case, we do not expect that Stark broadening
will be important for the Mg II resonance lines. 
Fleurier, Sahal-Brechot \& Chapelle (1977) determined experimentally that
the Stark broadening of these lines varied with core width between 
0.05-0.10\AA, and wings extending to $\sim$ 0.1-0.2\AA, for
temperatures of $1-3 \times 10^{4}$K and an electron density
of $N_e = 10^{17}$ cm$^{-3}$, much higher than expected in
the M dwarf atmosphere.  Figure 9 illustrates a simple Gaussian
fit to the underlying broad component in F9b yielding a 
redshift of 0.5\AA\ $\pm$ 0.08\AA\ (53 $\pm$ 8 km sec$^{-1}$)
and FWHM of 2.36\AA\ $\pm$ 0.58\AA\ (250 $\pm$ 60 km sec$^{-1}$).   If 
interpreted as velocities, the FWHM would suggest
turbulent emitting material with typical speeds $>$ 100
km sec$^{-1}$ within a bulk condensation moving
downward at roughly 50 km sec$^{-1}$.  Both the bulk and turbulent velocities
exceed the sound speed in the chromosphere, which is 
roughly 10-20 km sec$^{-1}$.  Adopting a chromospheric mass density of 
$1 \times 10^{-10}$ g cm$^{-3}$
during the flare (Allred et al. 2006) and a Mg II 
emitting region of height 300 km (see Figures 10 and 11 below) 
along with a flare area of 0.5\% of the visible stellar surface, 
the kinetic energy implied by chromospheric material moving 
at such large velocities far exceeds the energy radiated by the flare,
by several orders of magnitude.

Wood et al. (1996) found that the Mg II h and k lines in HR 1099,
an active RS CVn binary system, showed a broad component during
quiescence that looked similar to the broad components observed
in the C IV transition region resonance lines, with FWHM $\sim$
170 km sec$^{-1}$.  They were skeptical, as we are, that the
dense, optically thick chromosphere could respond to explosive
microflaring events to drive mass motions at these velocities,
and carried out detailed modelling that led them to suggest
that opacity effects in the line wings might account for
the observed broadening.  Again however, these effects were
associated with the star in quiescence, not in the midst of
an obvious strong flare.

\section{Comparsion to Flare Models}

Recently, Allred et al. (2005, 2006) produced models of solar and
stellar flares aimed
at understanding the chromospheric and transition region emission.
The models use solutions to the 1-D equations of radiative hydrodynamics,
including non-LTE radiative transfer in H, He and Ca II, with flare
heating provided by an electron beam.  The results include
predictions of atmospheric structure (temperature, density profiles),
velocity and line profiles at many time steps during an episode of
flare heating.
We used the results of their preflare
and F10 models to investigate the Mg II k emission during the YZ CMi flares.
The F10 model used here represents an average mid-flare atmosphere
near a time step of $\sim$ 85 sec; see Allred et al. (2006).
Note that the F10 model refers
to electron beam heating of 10$^{10}$ ergs sec$^{-1}$ cm$^{-2}$ 
during the model flare and is
not to be confused with our observed F10 flare presented in Section 2 above.

Since the Allred et al. models do not predict Mg II emission line
profiles, 
the preflare and F10 models were further analyzed with 
the `RH' non-LTE radiative 
transfer code described in Uitenbroek (2001). The RH code is based 
on the Multi-level 
Accelerated Lambda (MALI) formalism of Rybicki \& Hummer (1991, 1992), 
which allows both bound-bound and bound-free radiative transitions to 
overlap in wavelength. It also includes the effects of partial 
redistribution for strong bound-bound transitions such as Mg II h and k. 

Figure 10 shows the temperature and density (electron, hydrogen) structure
of the preflare atmosphere, together with the resulting Mg II k line
profile and the contribution function.  Figure 11 is a similar plot
for the model F10 flaring atmosphere.  
The contribution function indicates the atmospheric height where the
observed emission of interest is produced.  It is defined as the integrand
in the equation for the emergent intensity:
\begin{equation}
I_\nu(0) =\frac{1}{\mu} \int_{z} S_\nu \,
\chi_\nu \, e^{-\tau_\nu/\mu} dz
\end{equation}
where $I_\nu(0)$ is the emergent intensity at frequency $\nu$,
$\mu$ is the cosine of the angle between the vertical and the emergent ray
(we show the results for $\mu$ = 1, i.e. a ray that emerges vertically),
$S_\nu$ is the source function, $\chi_\nu$ and $\tau_\nu$ are 
the monochromatic opacity and optical depth respectively, 
and $z$ is the atmospheric height.

As shown in the figures,
the Mg II k line is formed in the upper chromosphere over a temperature
range from 8000K (line wings) to 20,000K (line core).  
At 20,000K the density is lower, hence there is relatively less 
emitting material.  This lower emission in the line core
causes the central reversal in the line profile (see Figure 10).
In the flaring atmosphere, there is an outward 
directed velocity where the core is formed, producing an 
asymmetric profile (see Figure 11) with a blue-shifted core.
The contribution function (center panel of Figure 11) also indicates that the 
red wing is formed in a lower, denser region than the blue wing. Therefore,
more flux is produced in the red wing.

We scaled the model flare line 
profile by 0.5\%, subtracted it from the pre-flare line profile
and smoothed it to the R=33,000 resolution of the data
to produce the Mg II k subtracted (flare-quiet) model line profile
shown in Figure 12.  Note that at this resolution, the
prominent central reversals in the model profiles are
much reduced, although the models still do not have
the smooth appearance of the observed line profiles in
Figure 4.  Nevertheless, the subtracted flare profile
in Figure 12 looks
quite similar to the observed profile during the impulsive 
phase of flare F9 (i.e. F9a, see Figure 8).  The velocity
shifts implied by the red-shifted emission are $\sim$ 4 km sec$^{-1}$
in the model profile, and $\sim$ 13.5 km sec$^{-1}$ in the observed
profile.

However, the models do
not reproduce the upward moving Mg II k material seen in the observed blue
enhancement flares (Figure 6), nor do they show the very broad
component in the decay phase of flare F9 (i.e. F9b, 
see Figure 9).  In fact, if the broadening is interpreted as
being due to emitting material moving with high velocity, the 
models never produce anything close
to the velocities required at the depth of the Mg II emitting
region.  

Alternatively, the very broad wings seen in flare F9 may be due
to enhanced redistribution of photons from the optically thick
Doppler core to the less thick radiative damping wings, due to
some process associated with the flare.
Comparing quiet and flare profiles in Figure 4 suggests that the
line core is saturated, even in the quiet atmosphere, so
redistribution is likely to be important.
The Allred et al. dynamical simulation includes the effects of
the electron beam on atmospheric
heating and the statistical equilibrium of H, but not Mg.
The RH code that we use to generate spectra from the Allred et al.
flare model includes a standard implementation of partial
redistribution.
A ``coherency fraction'' (see Uitenbroek 2001, eqn. 13) mixes
coherent scattering and complete redistribution in proportion to
the depopulation rate (due to radiation and inelastic
collisions) and elastic collision rate, respectively.
Typically, elastic collisions are due to van der Waals and
Stark broadening.
However, the RH code does not include the contribution to the
elastic collision rate due to interactions with the electron beam.

We speculate that the inclusion of the beam
collisions would enhance the collisional rates,
increasing the redistribution of line core photons into the wings
and possibly producing the far wing line emission seen in the data.
We plan to investigate this possibility in our next generation of
flare models.

\section{Summary}

We analyzed six orbits of HST/STIS near-ultraviolet spectroscopic data 
on YZ CMi and found that the star was flaring approximately 40\% of
the time.  The flares contributed $\sim$ 8\% of the energy observed
in the near-ultraviolet (2300-3050\AA) during the 224 total minutes
of exposure time.  Ten flaring and eight quiet intervals were identified,
and two of the flares (F2, F9) were large enough to divide into
sub-intervals to study the time evolution of the near-ultraviolet
emission during those flares.  The Mg II k emission line was examined
in detail and found to change significantly between different flares,
showing blue enhancements in some flares, red enhancements in others,
and symmetric enhancements in the wings or core in still others.
This indicates that Mg II emitting material may experience outward
or inward directed velocities (i.e. evaporation or condensation)
depending on the individual flare heating.  
The strongest Fe II line (UV1) showed similar behavior to the Mg II k
line.
Further, flare F9
showed a remarkable line broadening, which fit a Gaussian profile
corresponding to a velocity FWHM of 
$\sim$ 250 km sec$^{-1}$.  This velocity is highly supersonic, and if
the broadening is interpreted as due to either symmetric mass motions, or
turbulence, implies a kinetic energy during the flare that 
vastly exceeds the radiated energy (by several orders of magnitude).  
Overlapping emission due to explosive microflares, as has been
suggested to explain the broad components observed in optically
thin transition region emission lines, does not appear to be
a physically plausible explanation for the emission from
the Mg II lines which are formed in a dense, optically thick regime.

The data were compared to Mg II k line profiles generated using
the preflare and F10 flare models of Allred et al. (2006) together
with the RH code of Uitenbroek (2001).  The model results indicate
that the flare line profile has a red enhancement similar to that
seen in the observed F9a spectrum, with velocity of a few km sec$^{-1}$.
However, the models do not predict the blue enhancements seen in
some of the observed flares, nor does the Mg II emitting material
attain velocities anywhere near those required to match the broad
component of the Mg II k line seen in the F9b spectrum.  The
next step in the model calculations is to incorporate collisions
with the electron beam into the Mg II k partial redistribution 
calculation, which may have the effect of increasing the emission
in the radiative damping wings of the line, and could therefore
explain the observed line broadening.

SLH, LMW and JCA acknowledge the support of STScI grants HST-AR-10312,
HST-GO-10525,
and NSF grant AST-0205875.  The authors are grateful to Han Uitenbroek
for allowing us to use his excellent RH code and helping with
the implementation of the model atoms.  Useful discussions were
held with Rachel Osten, Phil Judge, Chris Johns-Krull, Andrew Becker
and John Bochanski.  We also thank the anonymous referee, whose 
substantive comments helped us to improve the paper.


\begin{deluxetable}{llllll}
\tablecaption{Flare and Quiet Time Intervals}
\tablenum{1}
\tablecolumns{6} 
\tablewidth{0pc} 
\tabletypesize{\footnotesize}
\tablehead{\colhead{Expo} & \colhead{t$_{Beg}$} & \colhead{t$_{End}$} & \colhead{Label} & \colhead{Rate} & \colhead{$\sigma$(Rate)} }
\startdata
1 & 0 & 22 & Q1 & 14.4 & 6.7 \\
 & 22 & 32 & F1 & 30.0 & 15.8 \\ \hline
2 & 0 & 5 & F2a & 62.0 & 70.9 \\
 & 5 & 10 & F2b & 38.7 & 9.7  \\
 & 10 & 15 & F2c & 17.9 & 5.1 \\
 & 15 & 40 & Q2 & 7.7 & 5.6 \\ \hline
3 & 0 & 17 & Q3 & 12.2 & 5.0 \\
 & 17 & 40 & F3 & 25.7 & 13.4 \\ \hline
4 & 1 & 4.5 & F4 & 30.1 & 31.1 \\
 & 4.5 & 15.5 & Q4 & 7.6 & 7.8 \\
 & 15.5 & 19 & F5 & 43.3 & 24.2 \\
 & 19 & 27.5 & Q5 & 7.0 & 7.7 \\
 & 27.5 & 30 & F6 & 34.9 & 6.1 \\
 & 30 & 37 & Q6 & 10.2 & 5.9 \\
 & 37 & 39 & F7 & 164 & 121 \\ \hline
5 & 0 & 1.5 & F8 & 44.2 & 3.8 \\
 & 1.5 & 37 & Q7 & 21.0 & 7.1 \\ \hline
6 & 0 & 8 & Q8 & 8.0 & 8.0 \\
 & 8 & 11.5 & F9a & 55.2 & 34.1 \\
 & 11.5 & 17 & F9b & 37.4 & 17.9\\
 & 17 & 25 & F9c & 8.4 & 4.4 \\
 & 25 & 36 & F10 & 26.9 & 7.2 \\
\enddata
\end{deluxetable}

\clearpage

\begin{deluxetable}{llll}
\tablecaption{Table 2: STIS Exposure Start Times}
\tablenum{2}
\tablecolumns{4} 
\tablewidth{0pc} 
\tablehead{\colhead{Expo} & \colhead{UT Date} & \colhead{UT Time} & \colhead{JD} \\ 
\colhead{} & \colhead{} & \colhead{} & \colhead{} } 
\startdata
1 & 2000 Feb 05 & 17:36:15 & 2451580.23351 \\
2 & 2000 Feb 05 & 18:56:14 & 2451580.28905 \\
3 & 2000 Feb 05 & 20:32:51 & 2451580.35615 \\
4 & 2000 Feb 05 & 22:09:29 & 2451580.42326 \\
5 & 2000 Feb 05 & 23:46:06 & 2451580.49035 \\
6 & 2000 Feb 06 & 01:22:43 & 2451580.55745 \\
\enddata
\end{deluxetable}

\clearpage

\begin{deluxetable}{rrr}
\tablecaption{Energy Budget in Lines and Continuum}
\tablenum{3}
\tablecolumns{3} 
\tablewidth{0pc} 
\tablehead{\colhead{Interval} & \colhead{Lines (ratio)} & \colhead{Continuum (ratio)} \\
\colhead{} & \multicolumn{2}{c}{\small{10$^{-13}$ ergs s$^{-1}$ cm$^{-2}$}} }
\startdata
 F1 & -0.45  (-0.09) & 5.50  (1.09)  \\
 F2a & 1.71  (0.10) & 14.85  (0.90) \\
 F2b & 4.12  (0.45) & 4.98  (0.55)  \\
 F2c & 0.42  (0.46) & 0.49  (0.54)  \\
 F3 & -0.32  (-0.17) & 2.20  (1.17)  \\
 F4 & 1.15  (0.09) & 12.36  (0.91)  \\
 F5 & 5.19  (0.41) & 7.57  (0.59)  \\
 F6 & 1.56  (0.15) & 9.11  (0.85)  \\
 F7 & 10.66  (0.22) & 38.24  (0.78)  \\
 F8 & 4.52  (0.56) & 3.61  (0.44)  \\
 F9a & 5.94  (0.33) & 11.84  (0.67)  \\
  F9b & 5.22  (0.51) & 4.97  (0.49)  \\
  F10 & 1.75  (0.30) & 4.02  (0.70)  \\
  Total F & 1.78  (0.23) & 6.10  (0.77)  \\
  Quiescent & 27.14  (0.65) & 14.35 (0.35)  \\
\enddata
\end{deluxetable}

\clearpage

\begin{deluxetable}{lrrrr} 
\tablecolumns{5} 
\tablewidth{0pc} 
\tablecaption{Narrow Component Flare Velocity Shifts} 
\tablehead{ 
\colhead{Flare} & \colhead{$\lambda_c$ \small{$\AA$}}   & \colhead{$\Delta\lambda_c$ \small{$\AA$}}    &
\colhead{V \small{(km/s)}} &  \colhead{$\Delta$V \small{(km/s)}}}
\startdata 
F1 &	2803.53	& 0.02	&    $-$4.65	&	2.14\\
F2a &	2803.53	& 0.04	&    $-$4.65	&	3.21\\
F3 &	2803.54	& 0.05	&    $-$3.58	&	5.35\\
F4 &	2803.70	& 0.03	&      13.52	&	3.21\\
F7 &	2803.56	& 0.04  &    $-$1.44	&	4.28\\
F9a&	2803.70	& 0.04	&      13.52	&	4.28\\
F10 &   2803.59 & 0.04 &        1.59	&       4.28\\

\enddata 
\end{deluxetable} 

\clearpage


\begin{figure}
\figurenum{1}
\epsscale{0.8}
\plotone{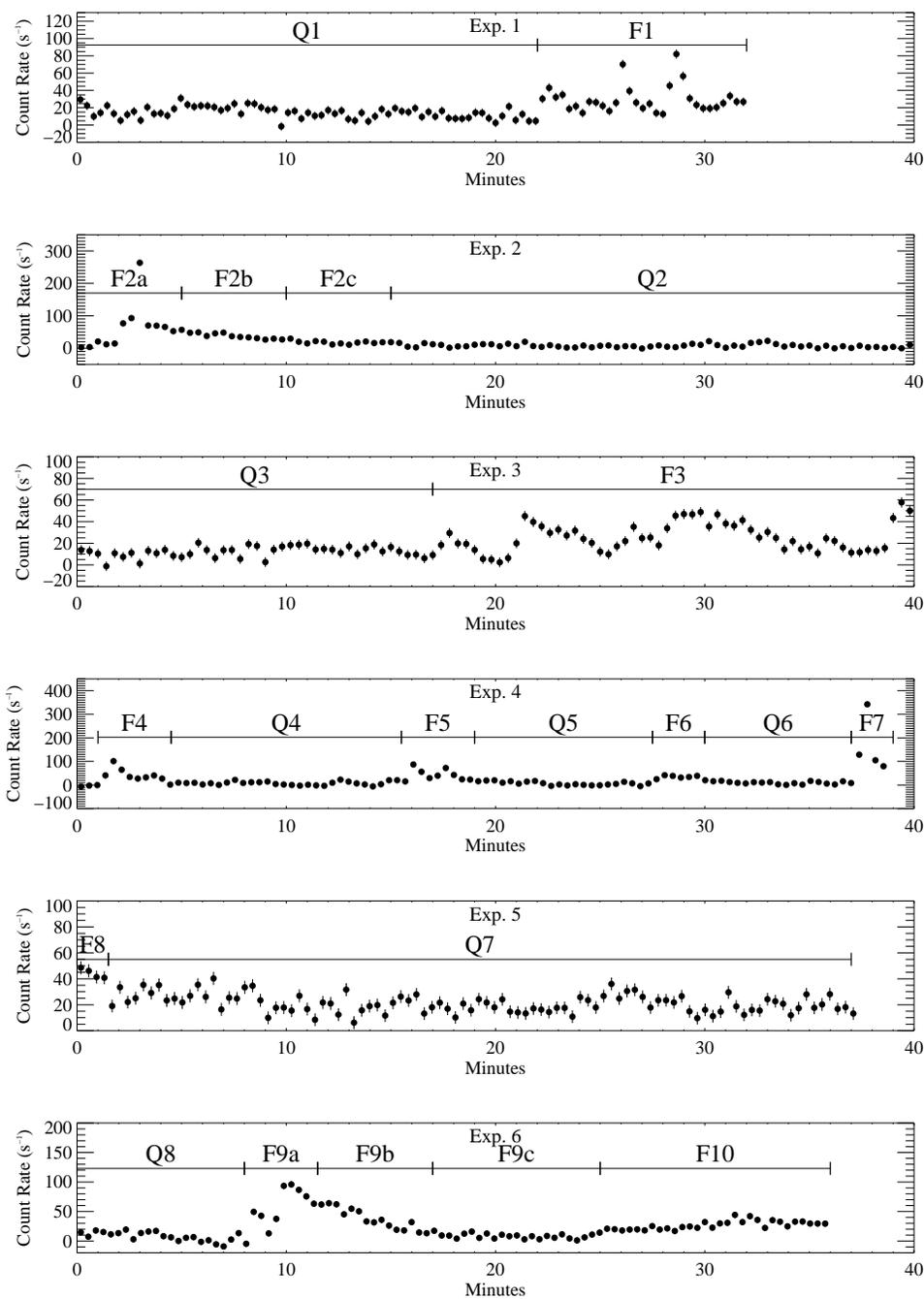}
\caption {Light curves showing the total count rate integrated over
the 2300-3050\AA\ STIS E230M bandpass during
each of the six exposures on YZ CMi.  See text for details about
the construction of the light curves.
The flaring (F1-F10) and quiet (Q1-Q8) intervals are indicated.}
\label{lightcurves}
\end{figure}

\clearpage

\begin{figure}
\figurenum{2}
\epsscale{0.8}
\plotone{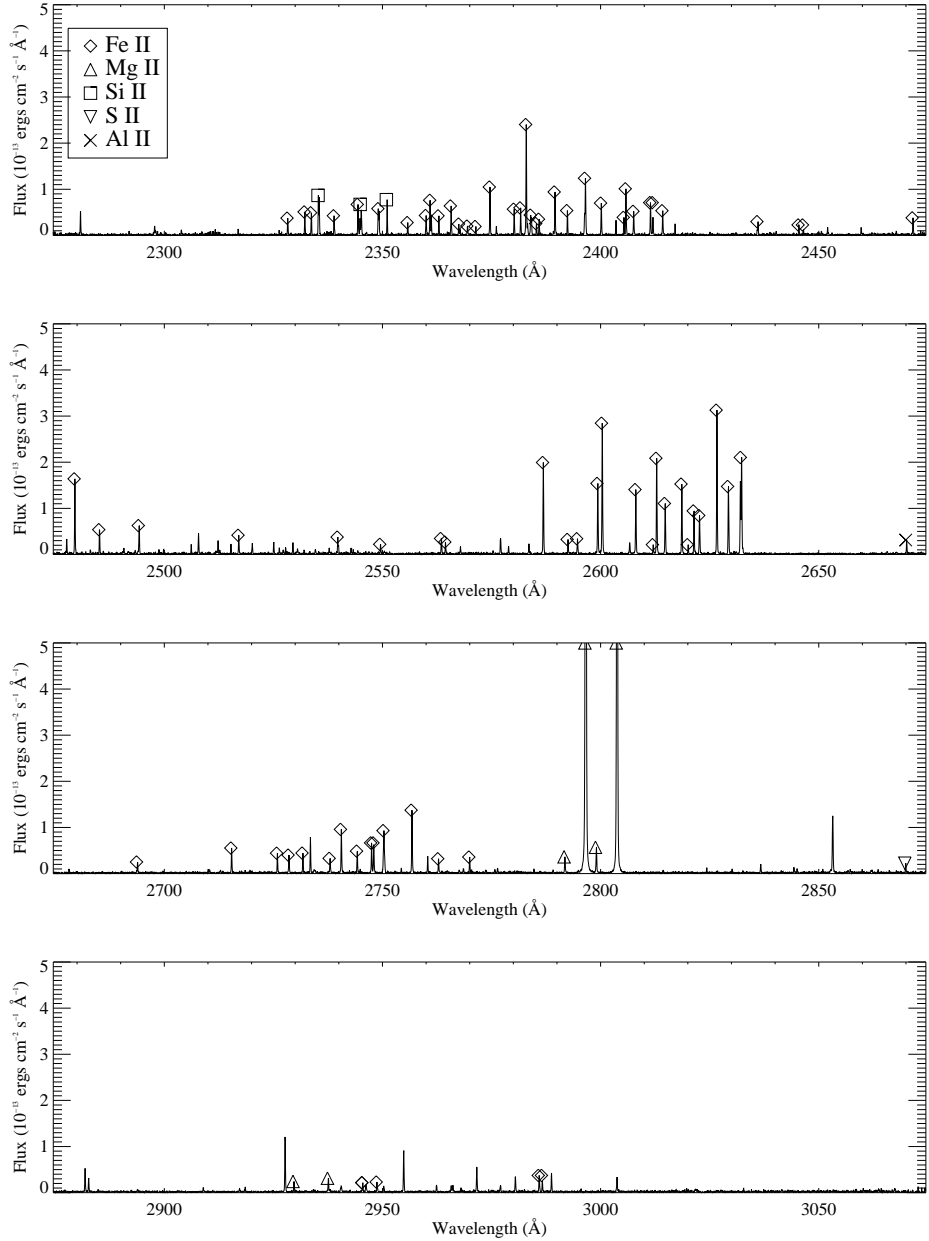}
\caption {All of the quiet intervals from Figure 1 have been
coadded to produce a high signal-to-noise quiet spectrum of
YZ CMi in the near-ultraviolet.  Emission lines identified in
the Chianti database are labeled.  The Mg II h and k lines 
extend off the figure, with peaks at $2.3 \times 10^{-12}$ and 
$1.8 \times 10^{-12}$ 
ergs sec$^{-1}$ cm$^{-2}$ 
respectively.}
\label{quiet}
\end{figure}

\clearpage

\begin{figure}
\figurenum{3}
\epsscale{0.8}
\plotone{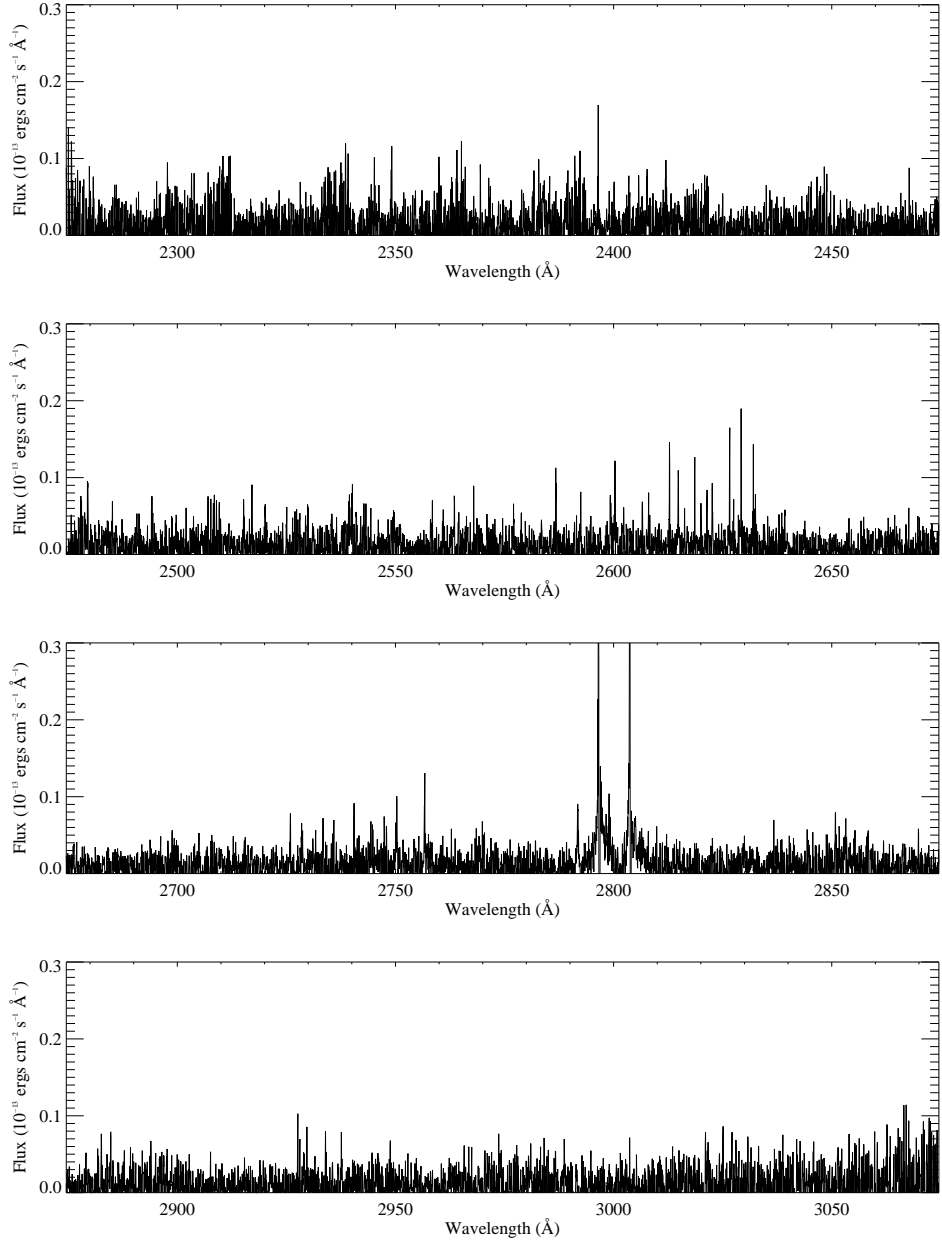}
\caption {The flare intervals from Figure 1 were coadded to produce a
flare spectrum, and the quiet spectrum from Figure 2 was subtracted
to produce this flare only spectrum.  The Mg II h and k
lines extend off the figure, with peaks at
$6.1 \times 10^{-14}$ and
$6.4 \times 10^{-14}$ 
ergs sec$^{-1}$ cm$^{-2}$ 
respectively.}
\label{fminusq}
\end{figure}

\clearpage

\begin{figure}
\figurenum{4}
\epsscale{0.8}
\plotone{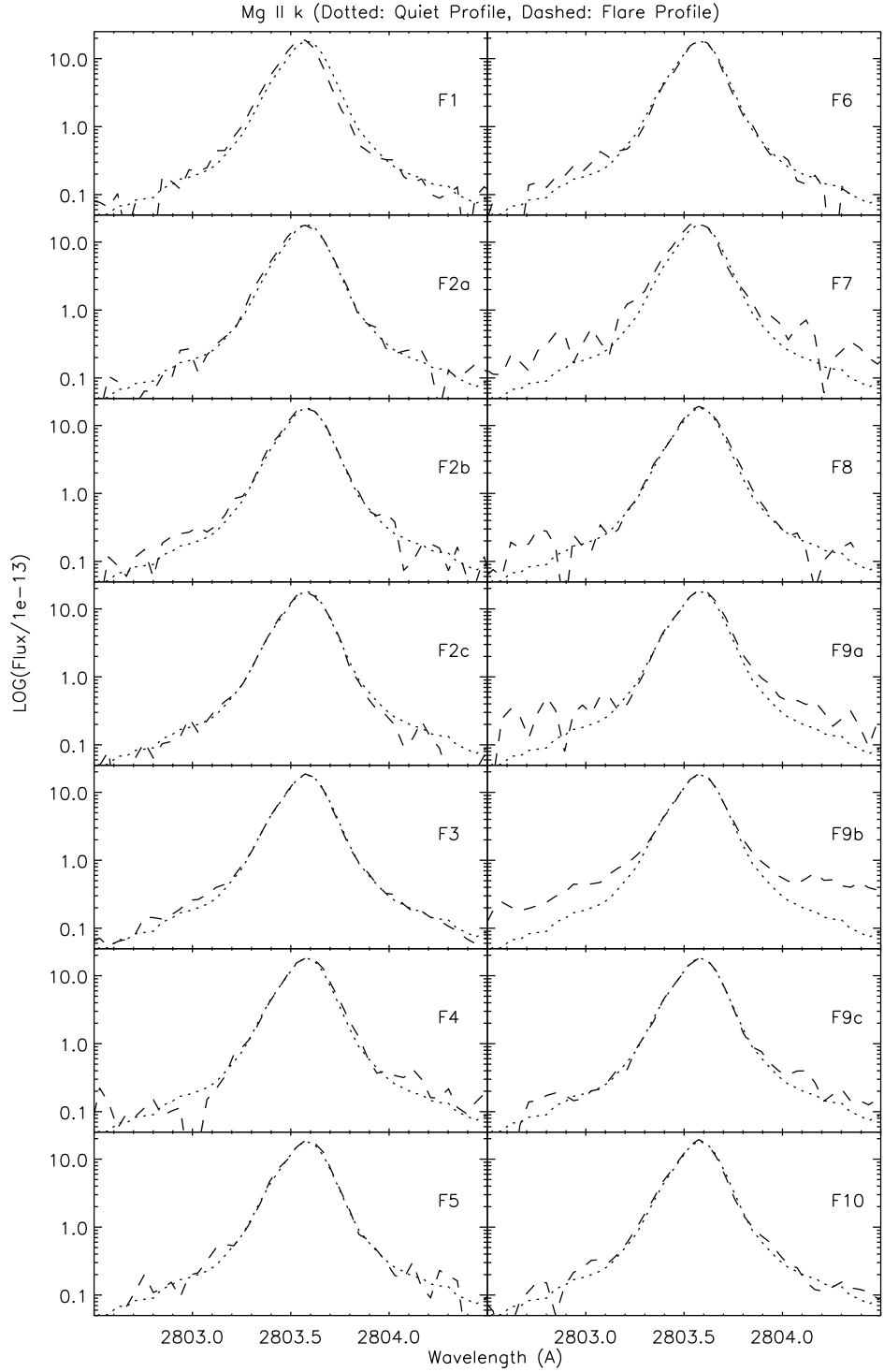}
\caption {The Mg II k line profiles are shown for each of the
flare intervals, including the three subdivisions of flares
F2 and F9.  The data are displayed on a log flux scale, so
that enhancements in the line wings during the flares
are evident.  The quiet profile is included for comparison in
each panel.}
\label{fig-mg2k}
\end{figure}

\clearpage

\begin{figure}
\figurenum{5}
\epsscale{0.8}
\plotone{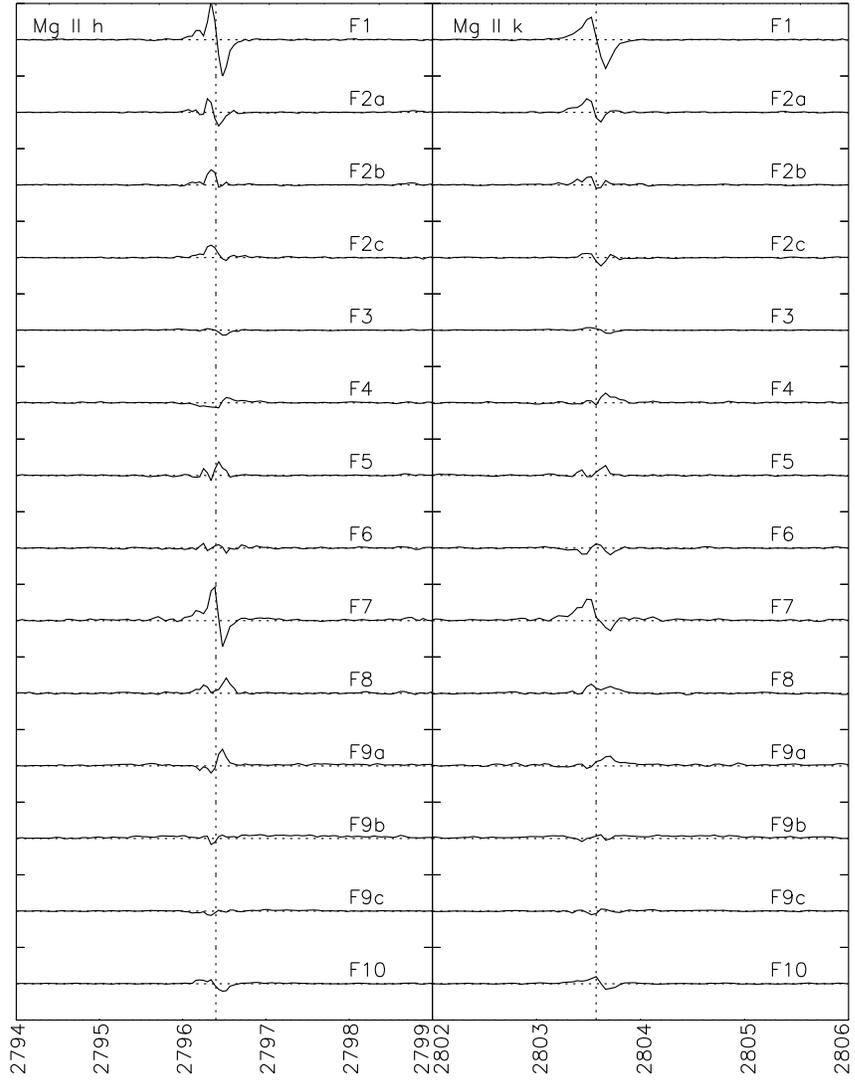}
\caption {The subtracted (flare-quiet) line profiles are shown
for Mg II h and k, indicating that the two lines behave very
similarly.  Each panel has the same scaling, so the relative
amount of flare emission during each flare interval can be
directly compared.  The impulsive flares in F1 and F7 and the
large F2 flare show the strongest response in the Mg II lines.}
\label{fig-mg2hksub}
\end{figure}

\clearpage

\begin{figure}
\figurenum{6}
\plottwo{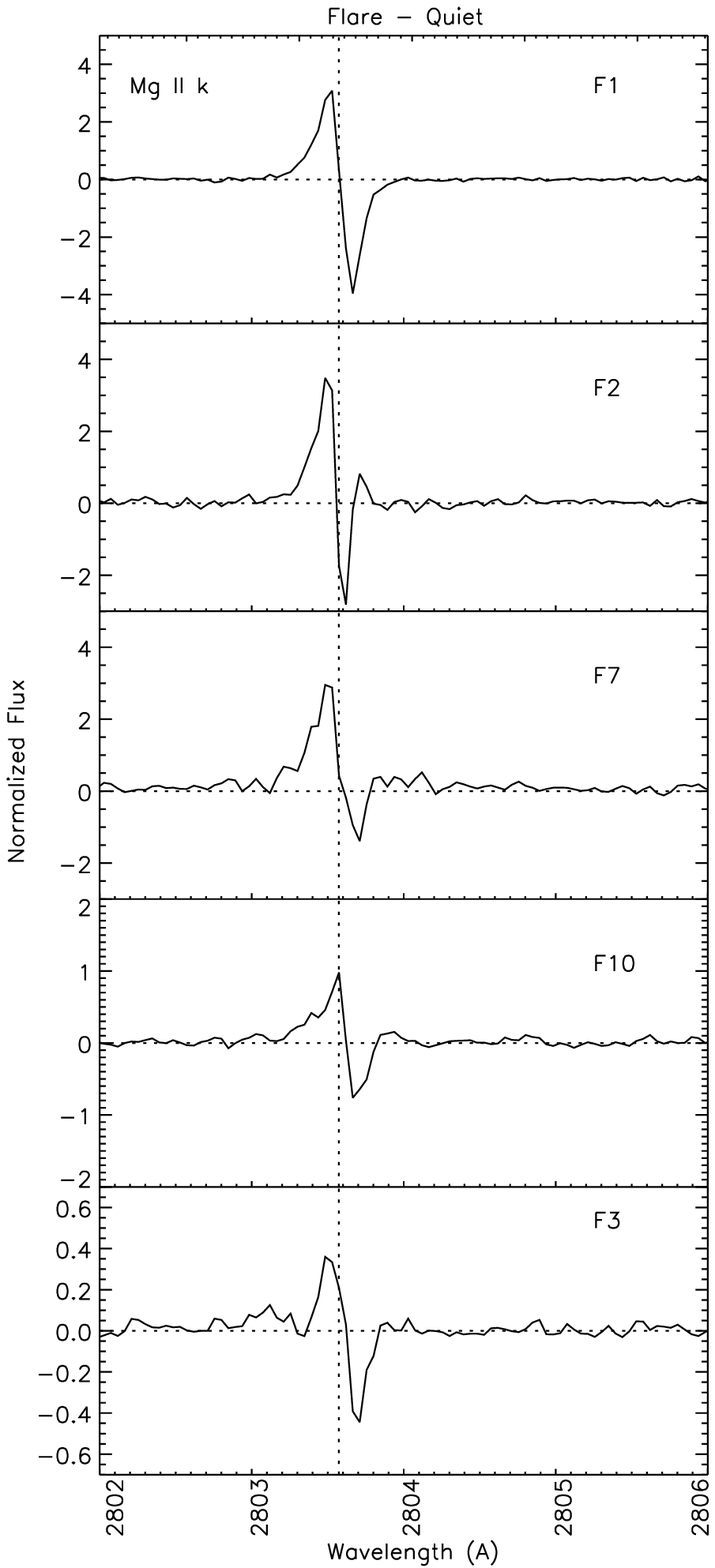}{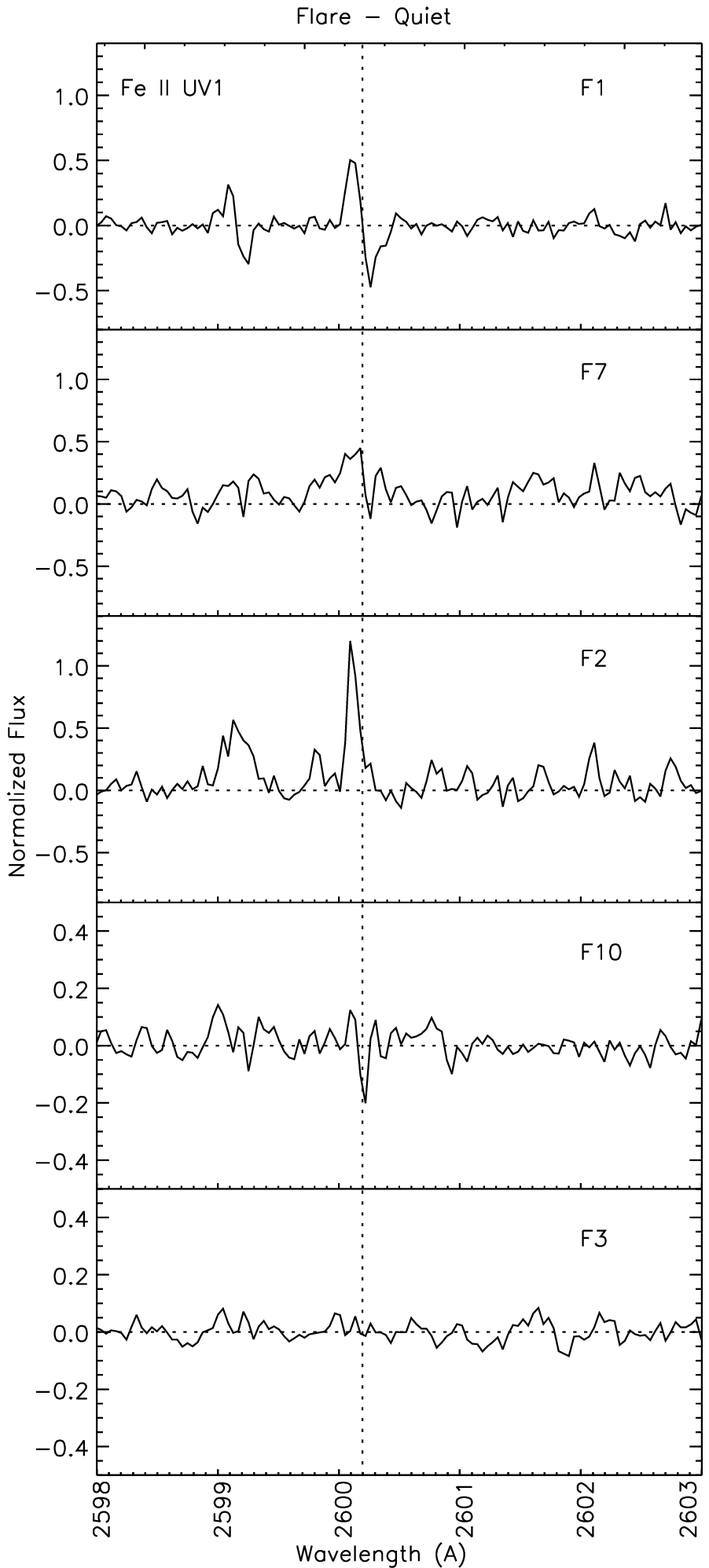}
\caption {The subtracted (flare-quiet) line profiles for the Mg II k
lines from Figure 5 that showed blue enhancements are illustrated 
in the left panels, in order from strongest to weakest emission
enhancement.  The Fe II UV1 line profiles for the same flare
intervals are given in the right panels.  Though noisier, the Fe II
data generally show the same blue enhancements as the Mg II data.}
\label{fig-bluemg2}
\end{figure}

\clearpage

\begin{figure}
\figurenum{7}
\epsscale{0.8}
\plottwo{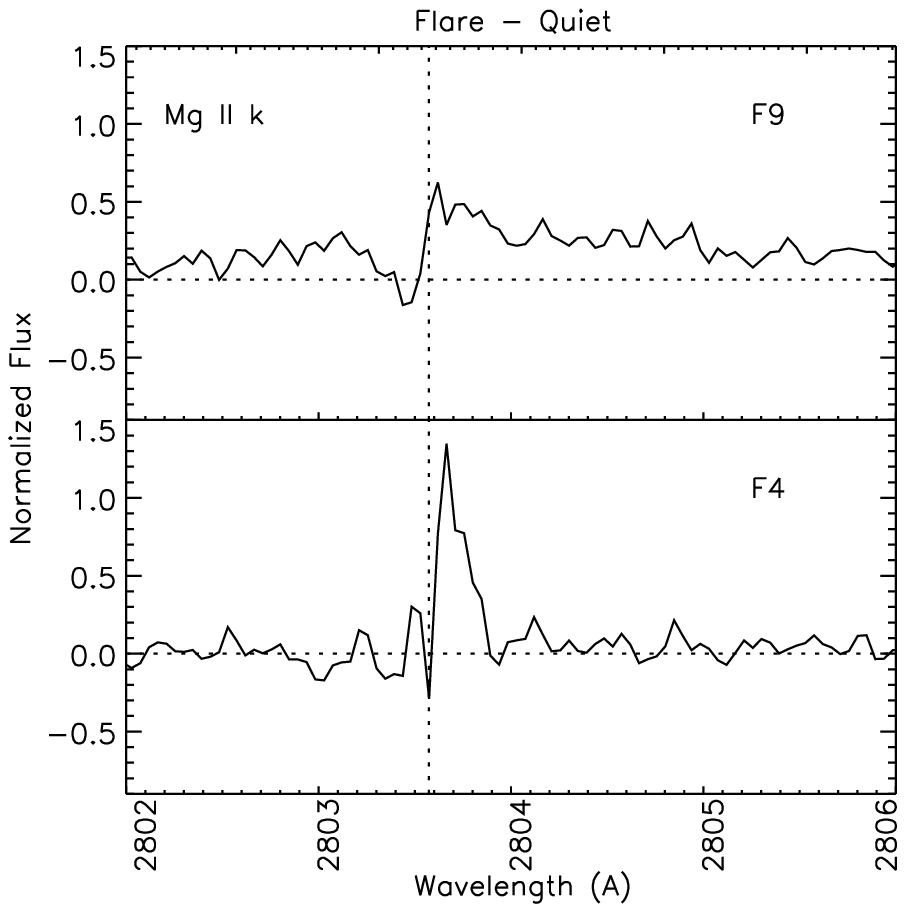}{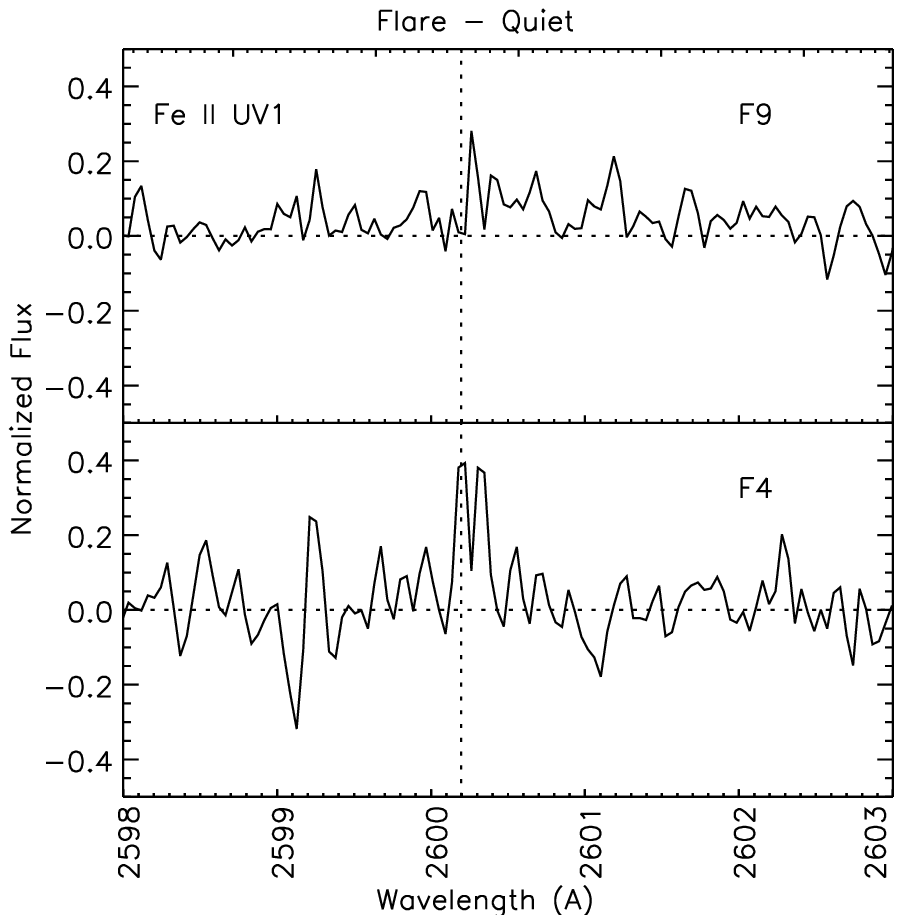}
\caption {Same as Figure 6, but for the flares that showed
red enhancements in Mg II k.}
\label{fig-redmg2}
\end{figure}

\clearpage

\begin{figure}
\figurenum{8}
\epsscale{0.8}
\plotone{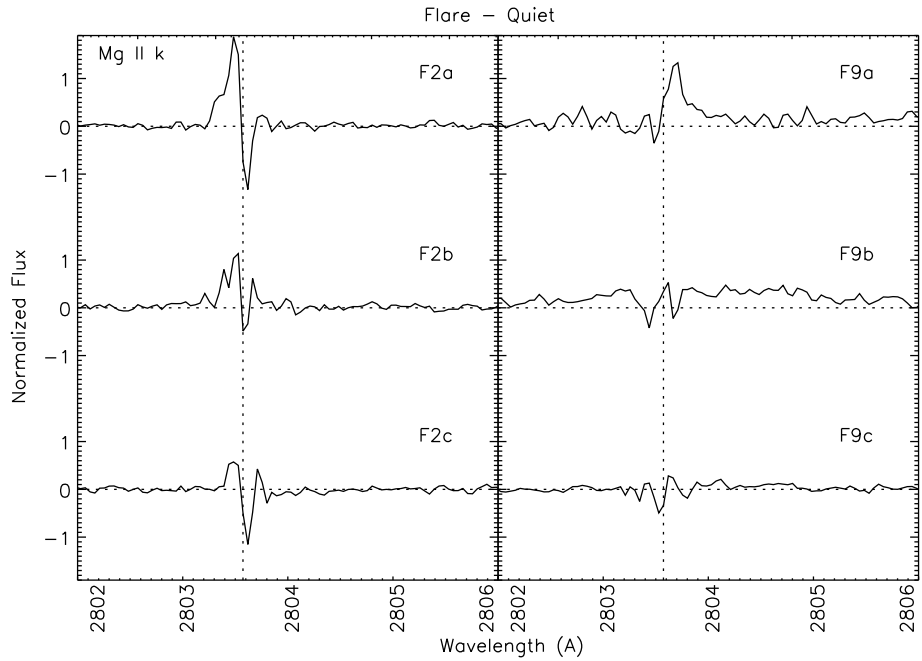}
\caption {The time evolution of the Mg II k subtracted 
(flare-quiet) line profiles is illustrated for intervals
F2abc and F9abc.  Flare F2 shows consistent behavior
throughout the impulsive and decay phases, while flare
F9 changes from a narrow, red enhanced line in F9a to
a symmetric but very broad line in F9b.  See text and
Figure 9 for further discussion of the broad line wings in F9b.}
\label{fig-subsets}
\end{figure}

\clearpage

\begin{figure}
\figurenum{9}
\epsscale{0.8}
\plotone{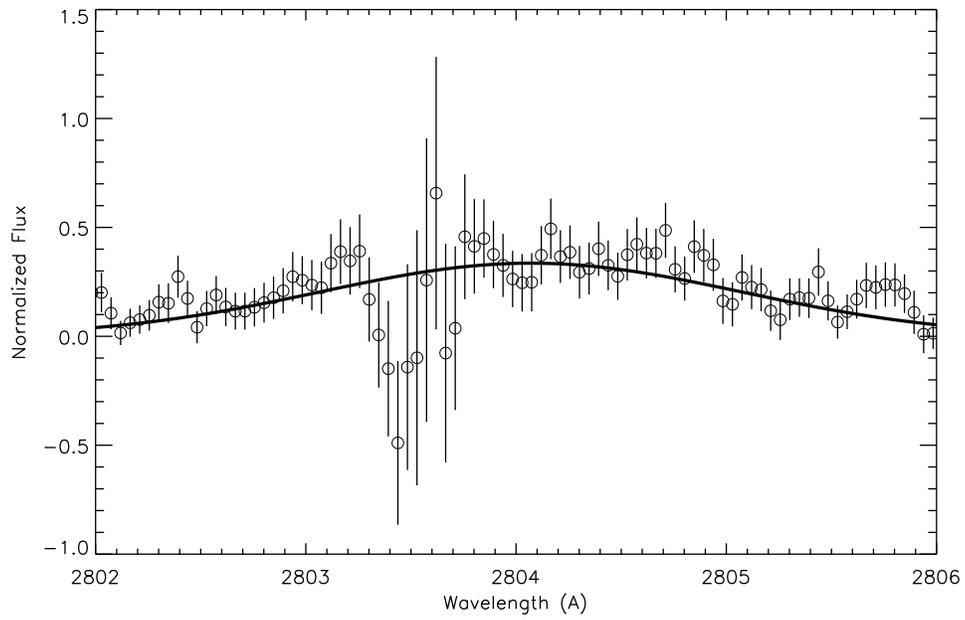}
\caption {The Gaussian fit to the broad component of
flare interval F9b is shown, together with the 
individual data points and uncertainties.  The
broad component is easily detected at the signal-to-noise
ratio of these data.  The fit parameters (given in
the text) imply a FWHM of 250 km/sec for this emission 
component.}
\label{fig-subsets}
\end{figure}

\clearpage

\begin{figure}
\figurenum{10}
\epsscale{0.8}
\plotone{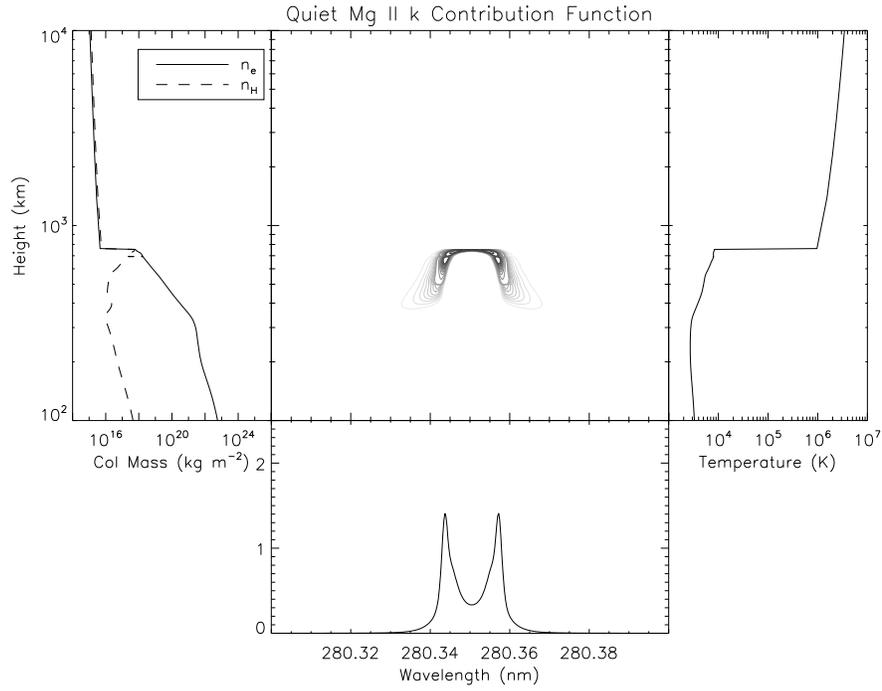}
\caption {The model temperature and density distributions (right and left
upper panels respectively), the Mg II k contribution function showing
where the Mg II k emission is being produced (middle upper panel), 
and the resulting Mg II k line profile (bottom panel) are illustrated
for the preflare (quiet) M dwarf model.}
\label{fig-modelpreflare}
\end{figure}

\clearpage

\begin{figure}
\figurenum{11}
\epsscale{0.8}
\plotone{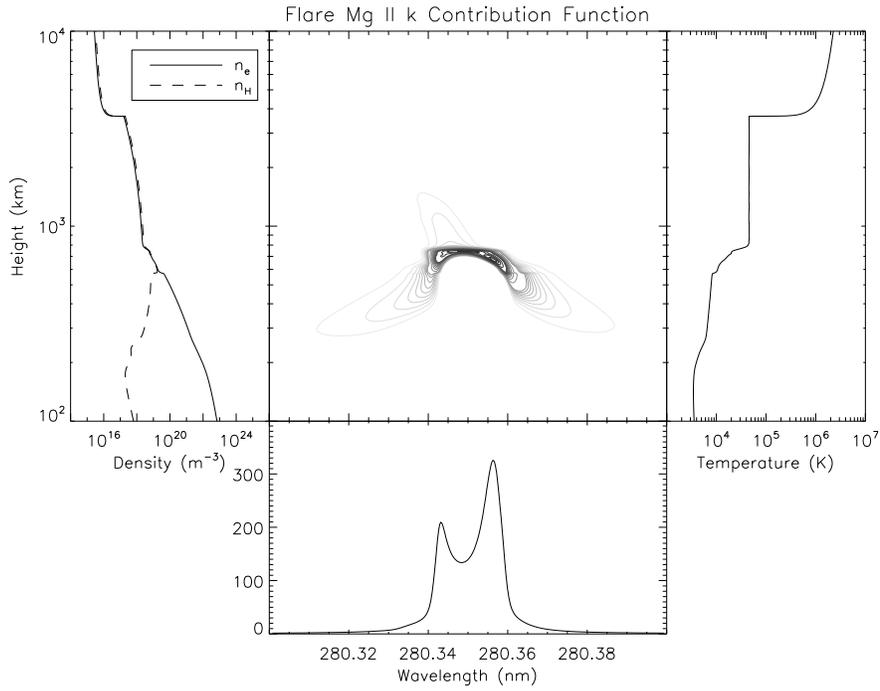}
\caption {Same as Figure 10 for the flare F10 model.  The flare
model clearly shows the effect of beam heating on the chromospheric
temperature and density distributions, and the contribution
function indicates that the Mg II emitting region is asymmetric
due to the velocity fields in the atmosphere. 
The line profile reflects this asymmetry.}
\label{fig-modelflare}
\end{figure}

\clearpage

\begin{figure}
\figurenum{12}
\epsscale{0.8}
\plotone{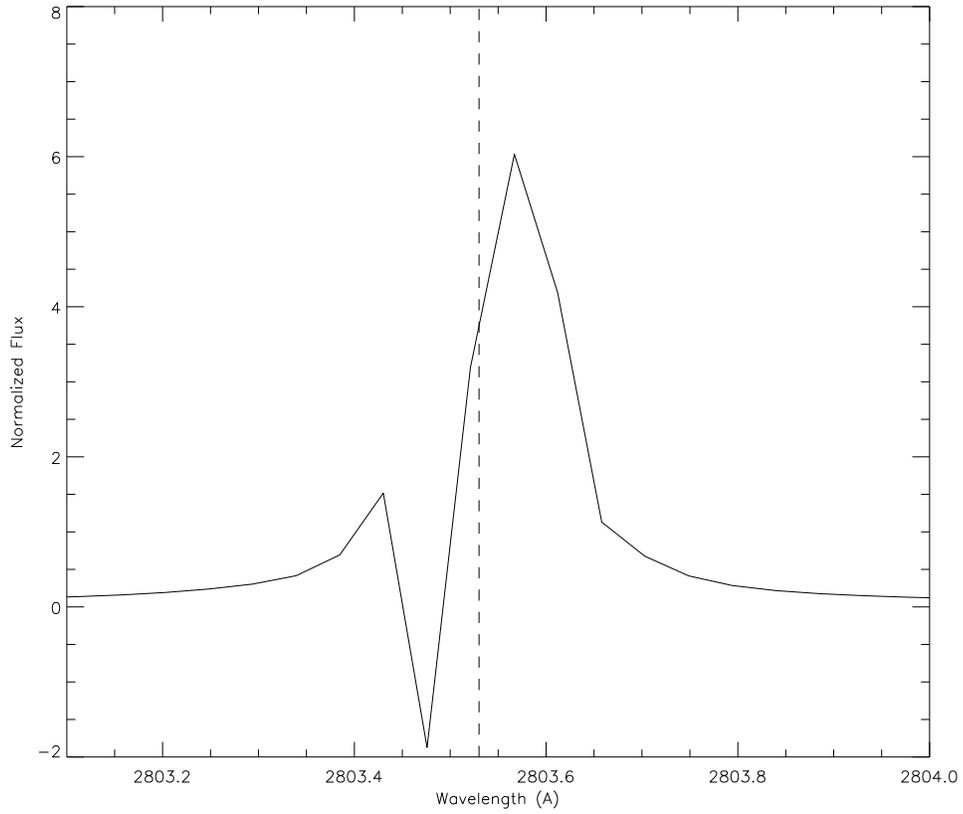}
\caption {The Mg II k subtracted (0.005 $\times$ model flare - model quiet) 
line profile
shows a red enhancement similar to that observed during flare F9a.
The flare area coverage was adjusted
so that the model flare emission is similar to the observed profile, giving a
value for the flare area of 0.5\% of the visible stellar surface.}
\label{fig-modelprofile}
\end{figure}

\end{document}